\documentclass[pra,10pt,superscriptaddress,footnoteinbib]{revtex4}
\usepackage{amsmath}
\usepackage{latexsym}
\usepackage{amssymb}
\usepackage{graphics,epstopdf}
\usepackage[colorlinks=true, citecolor=blue, urlcolor=blue ]{hyperref}
\usepackage{epsf,graphics,graphicx}
\usepackage{float}

  \usepackage{hyperref}
\newcommand{\n}{\noindent}

\newcommand{\beq}{\begin{equation}}
\newcommand{\eeq}{\end{equation}}

\begin{document}
	\title{Consequences of the thermal dependence of spin orbit coupling in semiconductors}
	\author{Debashree Chowdhury\footnote{Electronic address:{debashreephys@gmail.com\\~~~~~~~~~~
	debashreechowdury@hri.res.in}}} \affiliation{Department of Physics, Harish-Chandra Research institute, 
	Chhatnag Road, Jhusi, Allahabad, 
	U. P. 211019,India}


\begin{abstract}
\n
The $\vec{k}.\vec{p}$ perturbation theory in semiconductor modifies some spin related parameters of the semi-conducting system. Furthermore, renormalization of the Kane model parameters occurs when temperature appears in the scenario. In this paper, we have analysed the consequences of this renormalized Kane parameters on some spin transport issues. It is noteworthy to study that the temperature corrected scenario can open up a new direction towards the spin calorimetric applications in semiconductors.  
\end{abstract}

\maketitle
\n ~~~~~Keynotes: ESR frequency shift, thermal effects of SOC, spin torque

\section{Introduction}
Spintronics \cite{wolf,zutic,sh1} without magnetism is a new and fast developing field of research, where the spin properties of an electron within a semiconductor is investigated. In this scenario, the spin orbit coupling (SOC), which is the relativistic coupling between the spin and orbital angular momentum of electron, plays a crucial role. Besides, another important coupling is the Zeeman coupling, which arises because of the external magnetic field. The Zeeman term plays the key role for the explanation of the ESR frequency shift analysis.
Furthermore, the spin properties of the semiconductor is perturbed by the $\vec{ k} . \vec{ p}$ perturbation theory \cite{winkler} as the band structure of a semiconductor in the vicinity of the band edges can be very well described by the $ \vec{ k} . \vec{ p}$ method. It is possible to understand the characteristic features related to spin dynamics on the basis of $ \vec{ k} . \vec{ p}$ perturbation theory, where the interband mixing is taken care of\cite{kane}. As a result of $ \vec{ k} . \vec{ p}$ perturbation method we obtain the renormalization effects of spin orbit coupling (SOC) and the Zeeman coupling \cite{winkler}. These different spin couplings are modified due to the energy gap, spin orbit gap parameters \cite{winkler}, commonly known as Kane model parameters. It has been shown in \cite{bc}, the modified SOC term can provide useful information in the study of enhanced spin current, which is an important aspect in 
 spintronics device applications. The consideration of these renormalized  parameters \cite{winkler}, makes the theory of electron in semiconductors more accurate. Within a solid, the Zeeman strength or more precisely the electron $g$ factor is modified. In case of a free electron, the $g$ factor is nearly equal to 2.0023. The renormalized $g$ factor due to $\vec{ k} . \vec{ p}$ perturbation theory becomes $g + \delta g,$  where the $\delta g$ factor depends on the so called Kane model parameters\cite{kane}. There exists different theoretical and experimental models to measure the value of $g$ in different materials \cite{g1,g2,g3,g4,g5,g6,g7,g8,g9,g10}. But how the modified $g$ factor leads to ESR frequency shift is new and exciting scenario to look into.

However, the Kane model parameters are usually affected by temperature. Incorporation of these corrections in the spin related issues can lead to a new branch of spin physics, known as "spin caloritronics" \cite{scal}. Furthermore, due to inclusion of temperature we need to deal with modified SOC as well as modified electron $g$ factor. These motivate us to investigate the thermal effects on different spin related parameters. 

In this paper, the attention is paid on the frequency shift of the electron spin resonance (ESR) due to thermal effect and also on the development of the mechanical torque on spin precession. SOC is a very powerful ingredient to explain the physics behind spin Hall effect(SHE)\cite{SHE}. The modified SOC term \cite{bc}, due to Kane model parameters adjusts the spin current accordingly. For a particular choice of a semiconductor, it is possible to show the exact nature of variation of the spin current with temperature. Furthermore, as the SOC parameter is closely related to Berry curvature and Berry phase, these terms are highly affected by temperature as well.
  

The organization of the paper is as follows: in sec. II we build our model Hamiltonian considering the $\vec{ k} . \vec{ p}$ coupling between the bands $\Gamma_{6} $ (conduction band) and $\Gamma_{8} $ and $\Gamma_{7} $ (valance bands). For further reading jump to the appendix section. In sec. III, we have discussed the effect of temperature on different Kane model parameters. Also the thermal corrections on different spin related parameters are observed in Sec. IV. We have added a discussion in sec V, to compare the results with and without temperature corrections. The conclusion of the paper is presented in sec. VI. We have provided a appendix to describe clearly our notations and equations at the end of our conclusion section.
\section{The Model Hamiltonian}
The Pauli-Schr\"{o}dinger Hamiltonian with the effect of spin orbit coupling due to external electric and magnetic fields can be written as \cite{bc,cb,cbs}
\beq H = \frac{\hbar^{2}k^{2}}{2m} + qV(\vec{r}) + q\lambda\vec{\sigma} .(\vec{ k}\times \vec{ E}) + g\mu \vec{\sigma}.\vec{B} ,\label{Hk} \eeq
where the first and the second terms correspond to the kinetic term and the potential term respectively with $m$ as the free electron mass. The potential term contains the potential due to external electric field and the crystal potential. The third and forth terms are the spin orbit coupling term and Zeeman term appearing as a consequence of external electric and magnetic fields respectively. This free electron picture changes when we consider electrons within the semiconductors. One should incorporate the $ 8\times 8 $ Kane model to analyse the exact scenario, where we only consider the $\Gamma$ point splitting. Thus the scenario well describes the direct band semiconductor, where the minima of conduction band lies in the same line with the maxima of the valance band\cite{winkler}. We are interested in the Hamiltonian due to the $\vec{k}.\vec{p}$ perturbation, the detailed derivation of which is presented in the appendix. After some simple computation, we have arrived at the following 
 Hamiltonian \cite{winkler,bc}\beq
H_{kp} = \frac{P^2}{3}\left(\frac{2}{E_{G}} + \frac{1}{E_{G} + \triangle_{0}}\right)\vec{k}^{2} + eV(\vec{r}) - \frac{P^2}{3}\left(\frac{1}{E_{G}} - \frac{1}{(E_{G} + \triangle_{0})}\right)\frac{ie}{\hbar}\vec{\sigma}.(\vec{k}\times \vec{k})\\ + e\frac{P^2}{3}\left(\frac{1}{E_{G}^{2}} - \frac{1}{(E_{G} + \triangle_{0})^{2}}\right)\vec{\sigma}.(\vec{k}\times \vec{E})
\label{kp},\eeq where
$V(\vec{r}) = V_{e}(\vec{r})+ V_{c}(r),$ is the total potential of the system which contains potential due to the external electric field $V_{e}(\vec{r})$ and crystal potential $V_{c}(r)$. Here $E_{G}=E_c-E_v$ denotes the energy gap between the conduction and valance band with $E_{c}$ and $E_{v}$ denote the energies at the conduction and valence band edges respectively. $ \triangle_{0}$ is the spin orbit gap, $P$ is the Kane momentum matrix element which couples $s$ like conduction bands with $p$ like valence bands. This Kane Momentum matrix element remains almost constant for group III -- V semiconductors, whereas $ \triangle_{0}$ and $E_{G}$ varies with materials.  The parameters P, $ \triangle_{0}$ and $E_{G}$ are known as the Kane model parameters \cite{kane}.

Now to find out the total Hamiltonian, we must add up the Hamiltonians (\ref{Hk}) and (\ref{kp}). The total Hamiltonian for the electron at the conduction band edges can now be written as \cite{winkler}(see appendix)
\beq H_{tot} = \frac{\hbar^{2}\vec{k}^{2}}{2m^*} + qV(\vec{r}) + q(\lambda + \delta \lambda)\vec{ \sigma} .(\vec{ k}\times \vec{ E}) + (1+\frac{\delta g}{2})\mu \vec{\sigma}.\vec{B} ,\label{Hkp } \eeq where
$\frac{1}{m^*} = \frac{1}{m} + \frac{2P^2}{3\hbar^{2}}\left(\frac{2}{E_{G}} + \frac{1}{E_{G} + \triangle_{0}}\right)$ is the effective mass and
$\vec{ E} = -\vec{ \nabla} V_{e}(\vec{r})$ is the external electric field and $\lambda = \frac{\hbar^{2}}{4m^{2}c^{2}}$ is the spin orbit coupling strength in vacuum. $\delta g$ and $\delta \lambda,$ which are the renormalization factors for Zeeman and SOC terms respectively and can be written as 
\begin{eqnarray}\label{lam}
\delta g &=& -\frac{4m}{\hbar^{2}}\frac{P^2}{3}\left(\frac{1}{E_{G}} - \frac{1}{E_{G} + \triangle_{0}}\right)\nonumber\\
\delta \lambda &=& + \frac{P^2}{3}\left(\frac{1}{E_{G}^{2}} - \frac{1}{(E_{G} + \triangle_{0})^{2}}\right)
\end{eqnarray}
Including all the above parameters the Hamiltonian of our system can be written as 
\beq H_{tot} = \frac{\hbar^{2}k^{2}}{2m^*} + qV + q\lambda_{eff}\vec{\sigma} .(\vec{ k}\times \vec{ E}) + (1+\frac{\delta g}{2})\mu \vec{\sigma}.\vec{B} ,\label{eff} \eeq
where $\lambda_{eff}=\lambda+\delta\lambda$ is the effective SOC. The Hamiltonian in (\ref{eff}) is our system Hamiltonian. In (\ref{eff}), the first and second terms represent the kinetic term and the potential energy term respectively whereas the third and forth terms are associated with the modified SOC and Zeeman terms.  The renormalization of the mass and the SOC indicates that when we consider the electron within a semiconductor, we must take care of these Kane model parameters in the Hamiltonian.

The renormazied SOC parameter $\lambda_{eff}$ certainly influence the spin dynamics in of electron\cite{bc}. Due to the interband mixing of the bands the SOC parameter is changed, which consequently effects the spin Hall current. But how this SOC parameter modifies with temperature is of our current interest. Before going any further, in section III we would like to discuss the effect of temperature on the Kane model parameters.

\section{Effect of temperature on Kane model parameters}
Recently, semiconductor spintronics is a topic of great recent interest. In this connection, understanding the role of temperature on different spin related issues is an important concept. In this section, we are curious about the thermal corrections on different Kane model parameters. The dependence of the band gap parameter on temperature can be obtained as \cite{varshni, review}
 
 \begin{eqnarray}
E_{G}(T) = E_{G}(0) - \frac{\alpha T^{2}}{T+\beta},
 \end{eqnarray} 
where $\alpha,$ $\beta$  are the Varshni parameters \cite{varshni, review} and (0) denotes the values at $T = 0.$  As far as spin orbit gap is concerned, its variation with temperature is not clear in the literature. As it is also a gap parameter we can write a similar form as that of the energy gap parameter as \cite{debas} $\Delta_{0}(T) = \Delta_{0}(0) - \frac{\alpha^{'} T^{2}}{T+\beta^{'}},$ with $\alpha^{'},$ and $\beta^{'}$ as the extra pair of Varshni parameters. Here, we have considered the spin orbit gap parameter to behave as same as the energy gap parameter. This may not be the case as these two have different origins. The spin orbit gap parameter changes very slightly with temperature. Thus one can consider $\alpha^{'}\sim \beta^{'}\sim 0$, i.e these constants are very small numbers. This effectively gives us $\Delta_{0}(T)\simeq\Delta_{0}(0).$ In addition to the temperature dependence of the gap parameters, we 
 should also consider the temperature correction to the momentum matrix element $P$ as well. It is well known that the momentum matrix element $P$ varies with the lattice constant $a$ as $P \approx \frac{1}{a(T)}.$ Here, we have neglected the effect of phonon induced fluctuation of the inter-atomic spacing \cite{Hubner 2006}, for the simplicity of the calculations. The temperature dependence of the lattice constant can be written by the following relation \cite{Adachi}
\beq a(T) = a\left[1+ \alpha_{th}(T - 300)\right],\eeq
where $\alpha_{th}$ is the linear thermal expansion coefficient and the its value corresponds to the associated semiconductor. The values of Varshni's parameters as well as of $\alpha_{th}$ for two direct gap semiconductors can be given by\\
\begin{center}
\begin{tabular}{|*{2}{c|}l|}
\hline
$\alpha K^{-2}$ & $\beta(K)$ & $\alpha_{th}(K^{-1})$ \\
\hline
GaP = 5.8 $\times $ 10$^{-4}$ & 387 & 4.65 $\times$ 10$^{-6}$  \\
\hline
InP = 4.5 $\times $ 10$^{-4}$ & 335 & 4.65 $\times$ 10$^{-6}$\\\hline
\end{tabular}\\
\end{center}

\vspace*{.5cm}

It is important to note here that the Varshni parameters are strongly  material depended. Our focus here is to calculate the Kane model parameters having thermal corrections. In view of that one can write the Kane parameters as  \cite{debas}
\begin{widetext}
 \begin{eqnarray}
\frac{1}{m^*}  &=&  \frac{1}{m} + \frac{2}{3a^{2}\hbar^{2}(1+ \alpha_{th}(T - 300))^{2}}\left(\frac{2}{E_{G}(0) - \frac{\alpha T^{2}}{T+\beta}} + \frac{1}{E_{G}(0) - \frac{\alpha T^{2}}{T+\beta} + \triangle_{0}} \right)\nonumber\\
\delta g &=& -\frac{4m}{\hbar^{2}}\frac{1}{3a^{2}\hbar^{2}(1+ \alpha_{th}(T - 300))^{2}}\left(\frac{1}{E_{G}(0) - \frac{\alpha T^{2}}{T+\beta}} - \frac{1}{E_{G}(0) - \frac{\alpha T^{2}}{T+\beta} + \triangle_{0}}\right)\nonumber\\
\delta \lambda &=& + \frac{1}{3a^{2}\hbar^{2}(1+ \alpha_{th}(T - 300))^{2}}\left(\frac{1}{\left(E_{G}(0) - \frac{\alpha T^{2}}{T+\beta}\right)^{2}} - \frac{1}{\left(E_{G}(0) - \frac{\alpha T^{2}}{T+\beta} + \triangle_{0}\right)^{2}}\right).
\end{eqnarray}
\end{widetext}
Here the modification of the Kane parameters are due to the combined action of $\vec{k}.\vec{p}$ parameters and temperature effects. It is evident that the Zeeman and SOC terms are modified as $\mu_{B}\frac{(1+ \delta g(T))}{2}(\vec{\sigma}.\vec{B})$ and $q\lambda_{eff}(T)\vec{\sigma} .(\vec{ k}\times \vec{ E}) $ respectively. In the next section, we would like to point out the effects of the modified Zeeman and SOC terms on different spin related parameters.
\section{effect of temperature modified Zeeman and SOC terms}
It is noteworthy to mention that when the inter-band mixing of the bands are considered, some changes of the Zeeman and SOC term can be observed. Our aim in this section is to understand the effect of this altered zeeman and SOC terms on the physics of spin. The modified Zeeman term produces a shift in frequency and also a spin torque due to spin precession. We would like to put forward the expression of the shift in the frequencies as well as that of the spin torque. Furthermore, the thermally modified SOC changes the spin current as well as spin conductivity moderately. In the following two subsections we are focused to address the issues one by one.
\subsection{Frequency shift due to temperature}
The goal of this subsection is to deal with the effects of the temperature modulated Zeeman coupling on electron spin resonance(ESR). The phenomenon of ESR is based on the fact that an electron  has a magnetic moment and it respond in the external magnetic field. When an external magnetic field is imposed on the system, the electron spin will align itself with the direction of this field. But when we consider the electron within a semiconductor, the Zeeman energy term is renormalized due to the renormalized electronic "g" factor \cite{Annals rep}. Additionally, temperature adds new features to this story. The temperature corrected Zeeman term in semiconductor can be written as,  
\begin{widetext}
\begin{eqnarray} \label{g} H_{Z} &=& \mu_{B}(1+ \frac{\delta g (T)}{2})(\vec{\sigma}.\vec{B})\nonumber\\
& =& L(T)(\vec{\sigma}.\vec{B}),
\end{eqnarray}
\end{widetext}
where $L(T)$ is defined in Appendix (eqn. (A8)).
This causes an energy difference in the two Zeeman splitted levels, which is
\beq E_{\uparrow \downarrow} = \pm \frac{1}{2}L(T) B.\eeq
The difference in energy between the two levels can be given as \beq \delta E = L(T) B.\eeq
This difference in energy causes modification in ESR frequency, which can be written as,
\beq \omega^{'}_{ESR}(T) = \frac{1}{h}L(T) B \eeq
This ratio of frequencies with and without the temperature corrected ESR frequency is
\beq \frac{\omega^{'}_{ESR}(T)}{\omega} =  \frac{(1+ \frac{\delta g (T)}{2})}{(1+ \frac{\delta g}{2})},\label{ratio}\eeq
where $\omega$ corresponds to the zero temperature ESR frequency. Thus incorporating the expression of $\delta g (T)$ and $\delta g$ in the equation (\ref{ratio}), we have
\beq \omega^{'}_{ESR}(T) = \omega \frac{L(T)}{1 -  \frac{2m}{\hbar^{2}}\frac{P^2(0)}{3}\left(\frac{1}{E_{G}(0)} - \frac{1}{E_{G}(0) + \triangle_{0}(0)}\right)}\eeq
Thus the frequency shift enhances in presence of temperature. The equation (14) denotes the required frequency of a photon to cause a transition.

Next, we are interested to investigate the sensitivity of the ESR experiment and verify whether the ratio of population of spin up and down states depends on temperature or not. The  net absorption or emission of photons is proportional to the number of
spins in the lower level and emission is proportional to the number of spins in the upper level. As a consequence, one can write the net absorption to be proportional to the difference $n_{\downarrow} - n_{\uparrow}.$ The ratio of populations at equilibrium is given by \beq n= \frac{n_{\uparrow}}{n_{\downarrow}}  = e^{(-\frac{\delta E}{KT})} = e^{-\frac{1}{KT} L(T) B}.\eeq
At ordinary temperatures and  magnetic fields, the exponent is very small and can be given by
\beq n= \frac{n_{\uparrow}}{n_{\downarrow}} = 1 - \frac{B}{KT}L(T) \eeq or one can write the above equation as 
\beq n= \frac{n_{\uparrow}}{n_{\downarrow}} = (1-\frac{B\mu_{B}}{KT}) + \frac{2mB\mu_{B}}{KT\hbar^{2}}\frac{1}{3a^{2}\hbar^{2}(1+ \alpha_{th}(T - 300))^{2}}\left(\frac{1}{E_{G}(0) - \frac{\alpha T^{2}}{T+\beta}} - \frac{1}{E_{G}(0) - \frac{\alpha T^{2}}{T+\beta} + \triangle_{0}}\right).\eeq
\begin{figure}\label{fig}
	\includegraphics[width=8.0 cm]{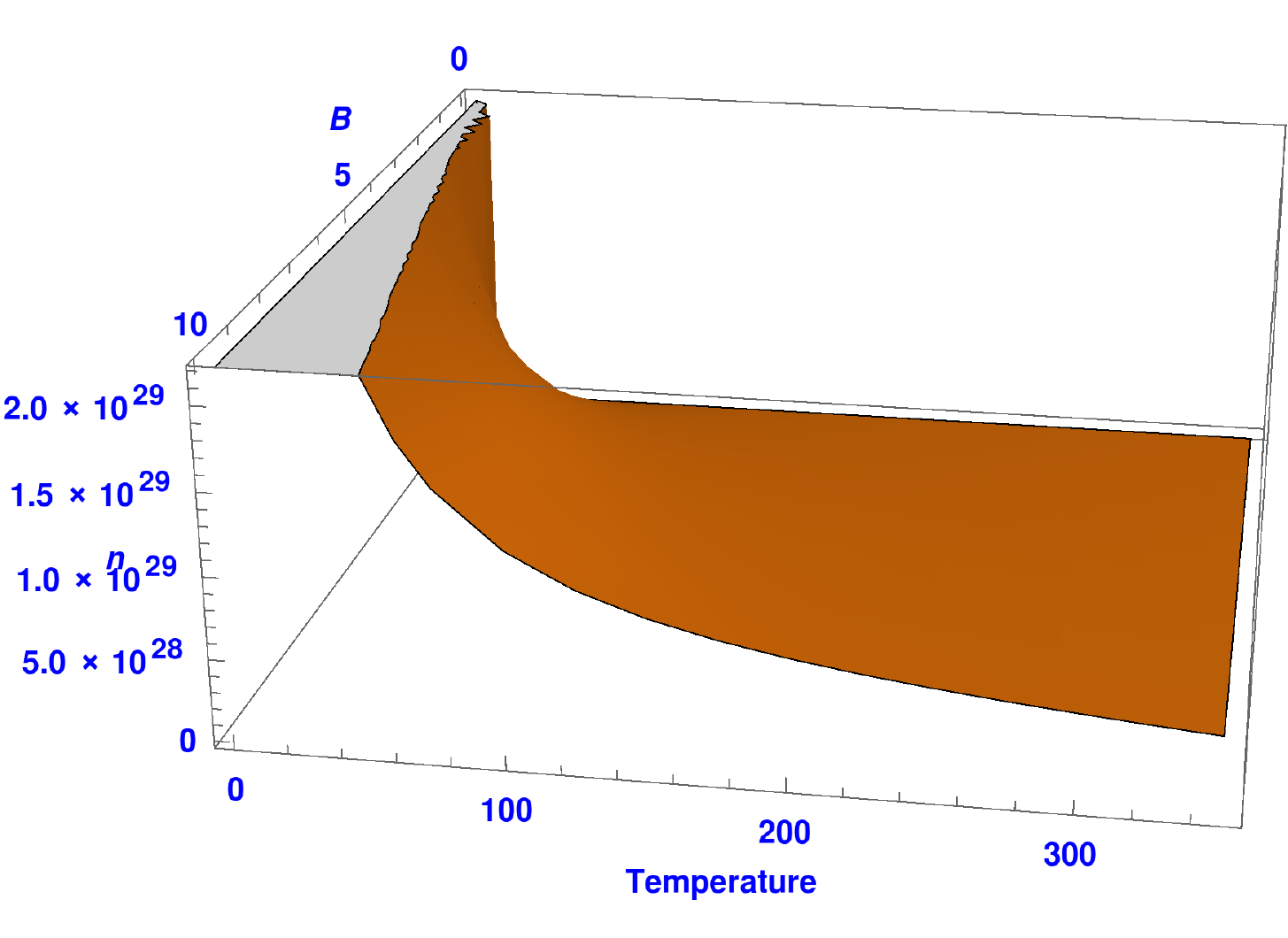}
	\hspace*{1cm}
\includegraphics[width=6.0 cm]{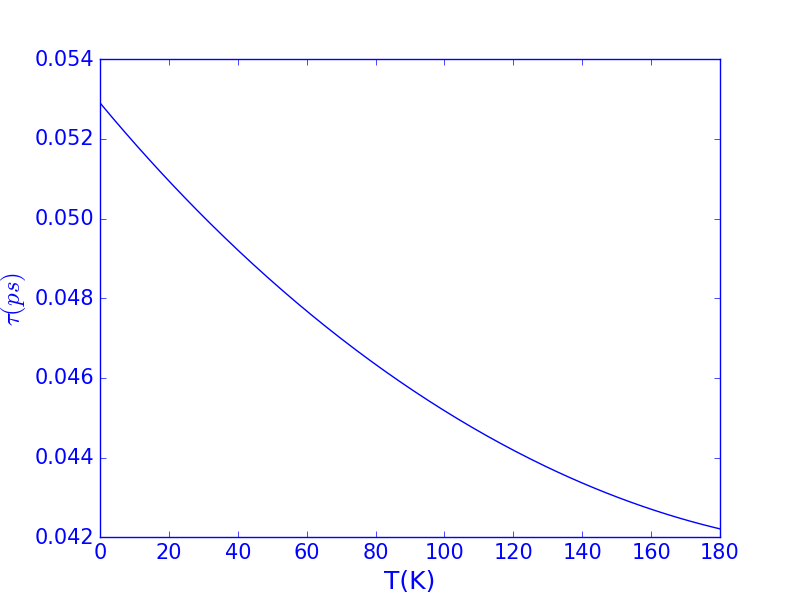}	
	\caption{ (Color online) Left: (a) Variation of ratio of population with magnetic field(tesla) and temperature(kelvin).
~~Right:(b) Variation of spin relaxation time with temperature. }
	
\end{figure}
This is the ratio of the two populations in the two spin splitted level, which depends on both temperature and the Kane model parameters. In FIG.1 (left) we have plotted the ratio of the two population with magnetic field and temperature. It shows that as we increase the temperature the ratio actually goes down. The result is opposite in case of increasing magnetic field. With increasing field, the ratio is increased. The left figure in  FIG 1 actually shows that when we increase the temperature the spin up population is decreased and spin down population will increase, that causes the ratio to fall down with temperature.
\subsection{Spin precession, corresponding torque and spin relaxation time}
When the electron is subjected to an external magnetic field, the spin of electron suffers a torque.
Considering the effect of SOC to be very small compared to the Zeeman term, one can write the equation for spin precession as,
\beq \dot{\vec{\sigma}} = \frac{1}{i\hbar}[\sigma, H],\eeq after incorporating Hamiltonian (1) in the above equation we have 
\begin{widetext}
\begin{eqnarray} 
\dot{\vec{\sigma}} &=& (1+ \frac{\delta g (T)}{2})\mu_{B}(\vec{\sigma}\times\vec{B})\nonumber\\
& = & L(T)(\vec{\sigma}\times\vec{B}). \end{eqnarray}
\end{widetext}
This gives the spin precession and it clearly depends on the temperature. Considering of the spin
relaxation and dephasing, the average spin of electrons with the magnetic torque is given by
\begin{eqnarray} 
\dot{\vec{S}} = L(T)(\vec{S}\times\vec{B}) - \frac{\vec{S}}{\tau_{s}}, \end{eqnarray} where $\vec{S} = \frac{1}{2}\hbar \vec{\sigma}$ is the is the spin angular momentum and $\tau_{s}$ is the spin relaxation time. This spin relaxation time depends on the SOC coupling parameter as
\beq \tau_{s}\propto \frac{1}{\alpha_{eff}} \propto \frac{1}{\xi(T)},\eeq
where $\xi(T)$ is defined in Appendix (eqn. (A9)).  
It can be argued from the above result that the spin relaxation time is influenced by the Kane model parameters and can be tuned with these band parameters. Not only this, we can also tune the spin relaxation time by using the thermal effects. Increase of spin relaxation time is very crucial for spintronics device applications. In FIG. 1:(right) we have plotted the variation of spin relaxation time with temperature for  InP semiconductor. The nature of the curve exactly matches with the experimental results described in \cite{bulk} for GaAs. The curve shows as we increase temperature the relaxation time is decreased. But to construct spintronic devices it is desirable to have longer spin relaxation time. From this point of view the conclusion is that, the incorporation of the thermal effects will not at all help us in designing a useful spintronics device.
\subsection{Spin current and conductivity}
In the above two subsections we have discussed mainly the effect of enhanced Zeeman term on spin related issues. Here we will examine the effect of the modifications of SOC term on spin dynamics. One of the important issue in this regard is to show the effect of the modified effective SOI on the spin current. In this subsection we consider that the SOC parameter is much larger than the Zeeman term. The relevant part of the Hamiltonian (for the positive energy solution) of spin $ \frac{1}{2}$ electron for zero external magnetic field can be expressed as
\begin{equation}\label{12344}
H =  \frac{\vec{p}^{2}}{2m^*} + eV(\vec{r})
- \lambda_{eff}\frac{e}{\hbar}\vec{\sigma}.(\vec{E}\times \vec{p})
\end{equation}

The semi-classical equation of motion of electron can be defined as
\beq \vec{F} = \frac{1}{i\hbar}\left[m^*(T)\vec{\dot{r}},H \right] + m^*(T)\frac{\partial\vec{\dot{r}}}{\partial t},\label{m3}\eeq with
$ \vec{\dot{r}}  = \frac{1}{i\hbar}[\vec{r}, H].$ Thus from (\ref{12344})
\beq \vec{\dot{r}} = \frac{\vec{ p}}{m^*(T)} -
\lambda_{eff}(T)\frac{e}{\hbar}\left(\vec{\sigma}\times \vec{ E}\right)\label{m},\eeq which is the expression of electron velocity. The second part of (\ref{m}) gives the anomalous velocity part, which depends on both the Kane model parameters and the temperature.
Incorporating the velocity term from (\ref{m3}) we have,
\begin{equation}\label{lor1}
\vec{F} = m^*(T)\ddot{\vec{r}} = -e\vec{ \nabla}V(\vec{ r})
+ \lambda_{eff}(T)\frac{em^{*}(T)}{\hbar}\dot{\vec{ r}}\times \vec{\nabla}\times
(\vec{\sigma}\times \vec{E}).
\end{equation}
The first and second terms of equation (\ref{lor1}) are the spin independent and spin dependent parts of the force and remind us about the Lorentz force of classical Hall physics. This force term can be called as "spin Lorentz force", with an effective magnetic field $\vec{\nabla}\times
(\vec{\sigma}\times \vec{E}).$ This spin Lorentz force is the main driving force for separating the spin up and spin down states. This separation is known as spin Hall effect in the literature. 
Explicitly, the magnetic field generated in the rest frame of the electron due to the effective SOI term, corresponds to a spin dependent gauge field. In our present case this gauge field also depends on temperature as
\begin{equation}\label{gauge}
\vec{A}(\vec{\sigma}, T) =   \lambda_{eff}(T)\frac{m^{*}(T)c}{\hbar}(\vec{\sigma}\times \vec{E})
\end{equation}
This spin dependent gauge can be shown to be correspond to the  modified $Aharonov-Casher(AC)$ \cite{casher} phase as 
\beq \phi_{AC} = \oint d\vec{r}. \vec{A}( \sigma, T) =  \lambda_{eff}(T)\frac{m^{*}(T)c}{\hbar}\oint d\vec{r}.  (\vec{\sigma}\times \vec{E}).\eeq 
Importantly, the AC phase is modified due to the combined action of temperature as well as band parameters. In this regard, it can be mentioned that the modified Zeeman term can't affect the $Aharonov-Bohm(AB)$ phase in the system as the external magnetic field remains unaffected by the temperature corrections. Thus the AB phase be unaltered as
\beq \phi_{AB} = \oint_{c} d\vec{r}.\vec{A}(r),\eeq
\begin{figure}\label{fi}
	\includegraphics[width=8.3 cm]{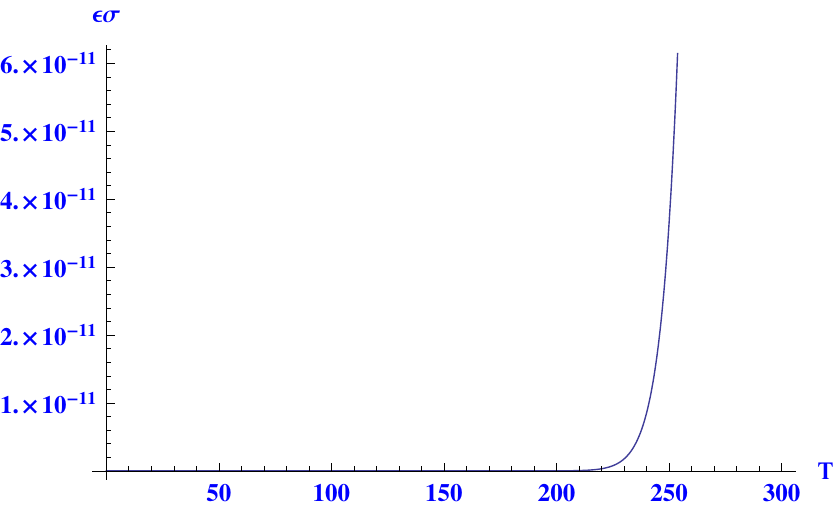}
		\hspace*{1cm}
		\includegraphics[width=8.3 cm]{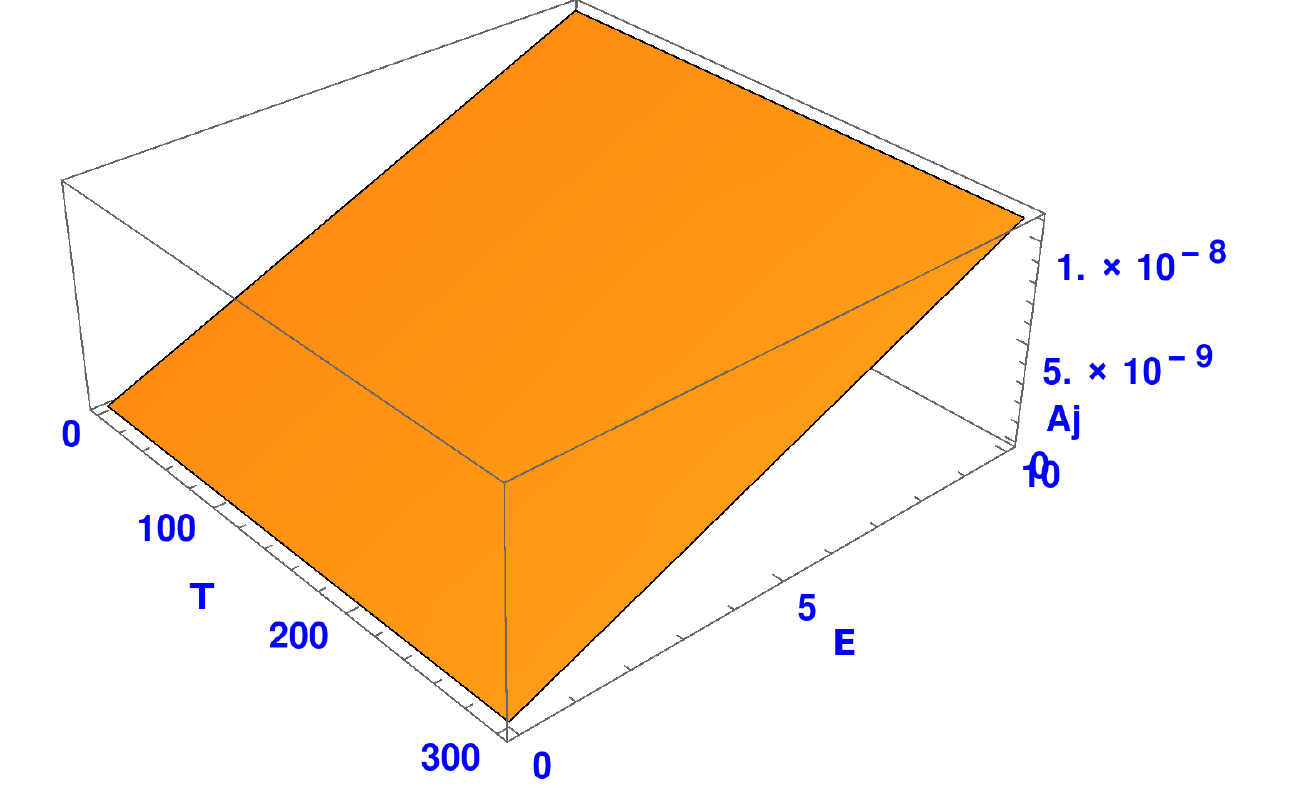}
	\caption{ (Color online)Left: Variation of conductivity with temperature with $\epsilon=\frac{1}{A}$. Right: Variation of spin current with temperature(K) and electric field(eV), with $A=(2\mu\rho\tau^{2})$}
\end{figure}
To derive the spin current and conductivity of this system we take resort to the method of averaging
\cite{n,cb} and consider that the spin dependent part of the force in (\ref{lor1}) is much smaller than the spin independent part. 
With the help of eqn. (\ref{gauge}), and neglecting $O(\vec{A}^{2}(\vec{\sigma})),$ the Hamiltonian (\ref{12344}) can be written as
\beq H = \frac{1}{2m^{*}}(\vec{p} -\frac{e}{c}\vec{A}(\vec{\sigma}))^{2} + eV(\vec{r}) \eeq
where $ V(\vec{r}) = V_{c}(\vec{r}) + V_{e}(\vec{r})$ where $V_{c}(r)$ is the crystal potential and  $V_{e}(\vec{r})$  is the external electric potential. 
The solution of
equation (\ref{lor1}) can be obtained as $\dot{ \vec{r}} = \dot{ \vec{r}}_{0} +
\dot{ \vec{r}}_{\vec{\sigma}}$\cite{n}. If the relaxation time $\tau$ is independent of $\vec{\sigma}$ and  for the constant electric field  $\vec{E}$, following \cite{n,cb} the spin current can be obtained. Here we choose a special kind of symmetry i.e cubic symmetry for which
one can write \cite{n, cb,cbs}
\beq \left\langle\frac{\partial^{2}V_{0}}{\partial r_{i}\partial r_{j}}\right\rangle = \mu \delta_{ij},\label{sym}\eeq
with $\mu$ being a system dependent constant.
The total spin current of this system with dual effect of temperature and $\vec{k}.\vec{p}$ perturbation can now be obtained as
\beq \vec{j}_{kp}(T) = e\left\langle\rho^{s}\vec{\dot{r}}\right\rangle = \vec{j}^{o}_{kp}(T) + \vec{j}_{kp}^{s}(T) \eeq
The charge component of this current is  \beq \vec{j}^{o }_{kp} = \frac{e^{2}\tau \rho}{m^*(T)}\vec{E} \label{jo} .\eeq
Introducing the density  matrix for the charge carriers as
\begin{equation}\label{dr}
\rho~^{s} = \frac{1}{2}\rho(1 + \vec{n}.\vec{\sigma}),
\end{equation}
where $\rho$ is the total charge concentration and $\vec{n}=\langle \vec{\sigma}\rangle$ is the spin polarization vector. One should note here that, the carrier concentration in semiconductor is also affected by temperature and for an intrinsic semiconductor one can write
\begin{figure}\label{fi}
	\includegraphics[width=8.0 cm]{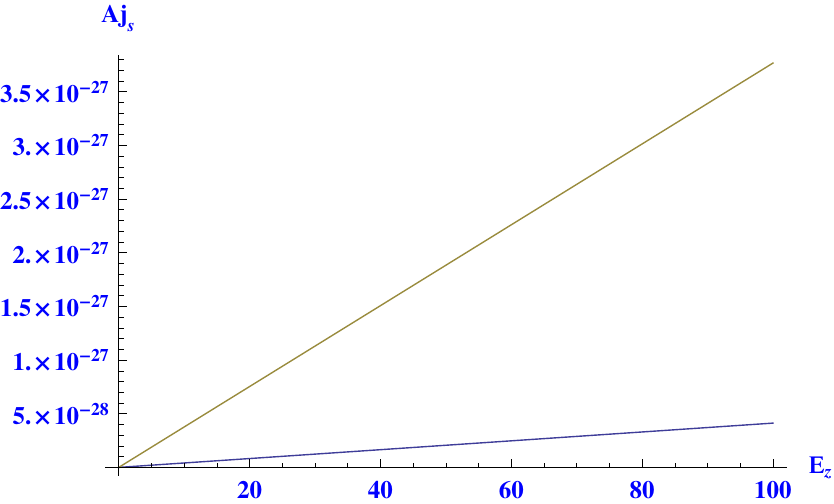}
	\caption{ (Color online) Variation of spin current with and without temperature corrections for semiconductor InP with $A= (2\tau^{2}\mu\rho)$.}
\end{figure}
\beq \rho(T) = 2(\frac{2\pi m^{*}KT}{\hbar^{2}})^{\frac{3}{2}}exp(\frac{-E_{G}(0)}{2KT})\label{h}.\eeq Incorporating eqn. (\ref{h}), we can write the spin current as (see appendix) 
  \begin{widetext}
\begin{eqnarray}\label{322}
\vec{j}_{kp}^{s}(T)  =\left(\frac{4e^{3}\tau^{2}\rho\mu}{\hbar}\right)(\frac{2\pi m^{*}KT}{\hbar^{2}})^{\frac{3}{2}}R(T)
exp(\frac{-E_{G}(0)}{2KT}) \left(\vec{n}\times \vec{E}\right)
\end{eqnarray}
\end{widetext}
It is evident from the equation (\ref{322}), that the spin current increases with the different values of the Kane model parameters.  It is shown in one of our earlier papers that the spin current depends on the material chosen. But in the present case, as all the Kane model parameters depend on temperature, the spin current also follow them. The spin conductivity then can be written as 
\begin{eqnarray}\label{324}
\sigma_{kp}^{s}(T)  &=& \left(\frac{2e^{3}\tau^{2}\rho\mu}{\hbar}\right)R(T)
\end{eqnarray}
 It is one of the major results of this paper.
For a particular semiconductor all the Varshini coefficients and the linear expansion coefficients are known. This helps us to give an exact variation of spin current and conductivity with temperature that can be measured in experiments. FIG. 2 left figure, shows the variation of spin conductivity with temperature, whereas the right figure elucidate the variation of spin current with temperature and electric field. This shows the increment of the spin current with temperature. In FIG:3 we have plotted the spin current with and without temperature corrections. This shows that the presence of temperature enhances the spin current, the magnitude of which is much higher in comparison to the non-thermal spin current. The main idea behind these results is that for a particular semiconductor the spin current and conductivity can change in a finite amount in presence of temperature. One should take care of the thermal effects in calculating spin related parameters in real systems.
\section{Discussion}
Here in this manuscript we have dealt with the correction factors of SOC and Zeeman terms due to temperature. For this we have used the Kane model and calculated the effective Hamiltonian. The main thing we need to point out here is that we obtain some modification in the ESR frequency shift and spin current due to temperature modified SOC and electron "g" factor. This quantities are different when we do not incorporate the temperature correction in SOC or electron g factor. Initially there exists comparatively large number of spin up electron. Importantly, when we consider the thermal modification of electron g factor, we can easily visualize that with increase of temperature the ratio decays and spin down electrons increase in number. Another interesting feature is we can write the value of the ratio with temperature correction as a function of the ratio without incorporating temperature correction due to electron "g" factor as
\beq n= n_{0}\frac{( 1 - \frac{B}{KT}L(T))}{(1- \frac{B}{KT}L(0)))}, \eeq
where $L(0)= \mu_{B}\left(1 -\frac{2m}{\hbar^{2}}\frac{P^2(0)}{3}\left(\frac{1}{E_{G}(0)} - \frac{1}{E_{G}(0) + \triangle_{0}(0)}\right) \right).$ Thus we have,
\beq \frac{n}{n_{0}}=\frac{( 1 - \frac{B}{KT}L(T))}{(1- \frac{B}{KT}L(0)))}\label{m2}.\eeq 

In case of the relaxation time when we do not include the temperature correction of SOC parameters, we have   
\beq \tau_{s}\propto \frac{1}{\alpha_{eff}} \propto \frac{1}{\lambda+\frac{P^2}{3}\left(\frac{1}{E_{G}^{2}} - \frac{1}{(E_{G} + \triangle_{0})^{2}}\right)},\eeq
which is much larger for this case as compared with eqn.(21). The spin relaxation time in temperature independent case is thus important from the prospective  of spintronics device application.

Next we have analysed the spin current and conductivity which are much larger in magnitude when we incorporate the temperature corrections. This is illustrated in figure 3, where it is evident that the spin current for temperature modified case is much larger than temperature independent case for a particular semiconductor InP. These are the important findings of the paper.
\section{Conclusions}
In this paper, based on the $\vec{k}.\vec{p}$ method, we  have investigated the thermal effects on spin related issues. The thermally renormalized SOC and Zeeman terms add some observable impacts to the story of spin transport. The present paper aims to investigate the thermal modulation of the spin precession, spin torque, spin current and conductivity parameters. The observations are as follows:\\

${\bullet}$ Firstly, the modulated Zeeman term causes a change in the ESR frequency. Apart from this, the modified Zeeman term is responsible for a temperature dependent torque in the system. This is very important issue in connection to spin precession and spin relaxation.\\

${\bullet}$ Secondly, the renormalized SOC is important in the sense that it has a deep impact on spintronics applications. The modified SOC in turn causes enhancement of the spin current of the system, which is a crucial concept for spintronics device applications. Through the cumulative effect of the Kane model parameters and temperature, the spin current is modified than that of the free case.\\

${\bullet}$ Finally, we would like to conclude by mentioning some of its future prospects. One can observe the effect of temperature on other parameters, for example the spin dependent gauges, which effectively modifies the Berry curvature and Berry phases. This is because of the fact that the incorporation of temperature changes the anomalous velocity term, which is directly related with the Berry curvature. 
Thus it may be noted, that our results include evaluation of some spin related quantities, which can also be experimentally verified. We hope our results will acquire sufficient interest to the experimental community. \\

\begin{center}
{\bf Appendix}:
\end{center}
The motion of electron's in a crystalline solid, is influenced by the energy bands $E_{n}(\vec{ k}),$ with band index $n$ and wave-vector $\vec{ k}$. SOC has a very profound effect on the energy band structure $E_{n}(k).$ Without SOC, there exists a $s$ like conduction band and p-like valance band. In case of the direct gap semiconductors, the SOC splits the topmost three fold degenerate valance band with total angular momentum $j = \frac{3}{2}$ (heavy hole and light hole) and $j = \frac{1}{2}$(split off holes), separated by a gap $\triangle_{0},$ which is known as the spin orbit gap. This helps us to realize the role of SOC on the orbital motion of electrons. This ensures that we have to deal with eight bands for a direct band semiconductor.
Thus for a $8$ band model, commonly known as Kane model \cite{kane}, which includes the $\vec{ k} . \vec{ p}$ coupling between the $\Gamma_{6}$ conduction band and $\Gamma_{8} $ and $\Gamma_{7} $ SOC split off valance bands,the Hamiltonian can be written as\cite{winkler} 
\begin{eqnarray}
	H_{8 \times 8}  =  \left( \begin{array}{ccr}
		H_{6c6c} & H_{6c8v} & H_{6c7v} \\
		H_{8v6c} & H_{8v8v} & H_{8v7v}\\
		H_{7v6c} & H_{7v8v} & H_{7v7v}
	\end{array} \right)
	~~=  \left( \begin{array}{ccr}
		(E_{c} + qV)I_2 & \sqrt{3}P\vec{ T} . \vec{ k} & -\frac{P}{\sqrt{3}}\vec{ \sigma} . \vec{ k}\nonumber\\
		\sqrt{3}P \vec{ T}^{\dag} . \vec{ k} & (E_{v} + eV)I_4  & 0 \\
		-\frac{P}{\sqrt{3}}\vec{ \sigma} . \vec{ k} & 0  & (E_{v} - \triangle_{0} + eV)I_{2}
	\end{array} \right),\label{a}\nonumber~~~~~~~~~~~~~~~~~~~~~~~~~(A1)             
\end{eqnarray}
where $T_{i}$s are given by
\begin{widetext}
	\begin{equation}
		T_{x}  = \frac{1}{3\sqrt{2}} \left( \begin{array}{ccrr}
			-\sqrt{3} & 0 & 1 & 0 \\
			0 & -1 & 0 & \sqrt{3}
		\end{array} \right),~~~~
		T_{y}  = -\frac{i}{3\sqrt{2}} \left( \begin{array}{ccrr}
			\sqrt{3} & 0 & 1 & 0 \\
			0 & 1 & 0 & \sqrt{3}
		\end{array} \right),
		T_{z}  = \frac{\sqrt{2}}{3} \left( \begin{array}{ccrr}
			0 & 1 & 0 & 0 \\
			0 & 0 & 1 & 0
		\end{array}\right)\nonumber~~~~~~~~~~~~~~~~~~~~~~~~~(A2)    
	\end{equation}
\end{widetext}
and $I_{2}, I_{4}$ are $2\times 2$
and $4\times 4$ unit matrices respectively.

The above $H_{8 \times 8}$  Hamiltonian can be written as a Pauli-Schr\"odinger equation for the conduction band electrons\cite{winkler}. For this we first start with the Schr\"odinger equation as, 

\begin{eqnarray}\label{6c}
E\Psi = H_{8\times 8} \Psi \nonumber~~~~~~~~~~~~~~~~~~~~~~~~~~~~~~~~~~~~~~~~~~~~~~~~~~~~~~~~~~~~~~~~~~~~~~~~~~~~~~~~~~~~~~~~~~~~~~~~~~~~~~~~~~~~~~~~~~~~~~~~~~~~~~~~(A3)
\end{eqnarray}
with $\Psi = (\psi_{6c}, \psi_{8v}, \psi_{7v})^{T},$ is the corresponding wavefunction. 

Incorporating eqn.(A1)  into (A3) , 
\begin{eqnarray}
& \Big[ &
\vec{T}\cdot \vec{k} \frac{3 P^{2}}{E_{G}} \Big(1+\frac{\tilde{E}-V}{E_{G}}  \Big)^{-1}  \vec{T}^{\dagger}\cdot \vec{k} 
+\vec{\sigma} \cdot \vec{k} \frac{P^{2}/3}{E_{G} +\Delta_{0}} \Big(1+\frac{\tilde{E}-V}{E_{G}+\Delta_{0}}  \Big)^{-1} \vec{\sigma} \cdot \vec{k}
\Big]\psi_{6c} \nonumber \\
&& = (\tilde{E}-V) \psi_{6c} \label{eq6c},\nonumber~~~~~~~~~~~~~~~~~~~~~~~~~~~~~~~~~~~~~~~~~~~~~~~~~~~~~~~~~~~~~~~~~~~~~~~~~~~~~~~~~~~~~~~~~~~~~~~~~~~~~~~~~~~~~~~~~~~~~~~~(A4)
\end{eqnarray}
where $\tilde{E} = E - E_{c}.$

From the rule of norm conservation, we have, 
\begin{eqnarray}
\int d^{3}x \, |\Psi|^{2} = \int d^{3}x \, | \tilde{\psi}_{6c} |^{2}.\label{norm}\nonumber~~~~~~~~~~~~~~~~~~~~~~~~~~~~~~~~~~~~~~~~~~~~~~~~~~~~~~~~~~~~~~~~~~~~~~~~~~~~~~~~~~~~~~~~~~~~~~~~~~~~~~~~~~~~~~~~(A5)
\end{eqnarray}
The l.h.s. of  eqn. (A5) is
\begin{eqnarray}
\int d^{3}x \, |\Psi|^{2} \approx
&& \int d^{3}x \, 
 1 + \frac{3P^{2} (\vec{T}\cdot \vec{k}) (\vec{T}^{\dagger}\cdot k)}{E_{G}^{2}} \nonumber\\
 &&+ \frac{P^{2} (\vec{\sigma}\cdot k) (\vec{\sigma} \cdot k)}{3(E_{G}+\Delta_{0})^{2}} \Big|\psi_{6c}\Big|^{2},\nonumber~~~~~~~~~~~~~~~~~~~~~~~~~~~~~~~~~~~~~~~~~~~~~~~~~~~~~~~~~~~~~~~~~~~~~~~~~~~~~~~~~~~~~~~~~~~~~~(A6)
\end{eqnarray}
where we neglect $(\tilde{E}-V)/E_{G}$ and $(\tilde{E}-V)/(E_{G} + \Delta_{0})$. 
Hence, we have
$\tilde{\psi}_{6c} = 
\Big[1+ N\Big] \psi_{6c} 
$ with
\begin{eqnarray}
N=
\frac{3P^{2} (\vec{T}\cdot \vec{k}) (\vec{T}^{\dagger}\cdot \vec {k})}{2E_{G}^{2}} 
+
\frac{P^{2} (\vec{\sigma}\cdot \vec{k}) (\vec{\sigma} \cdot \vec{k})}{6(E_{G}+\Delta_{0})^{2}}.\nonumber~~~~~~~~~~~~~~~~~~~~~~~~~~~~~~~~~~~~~~~~~~~~~~~~~~~~~~~~~~~~~~~~~~~~~~~~~~~~~~~~~~~~~~~~~~~(A7)
\end{eqnarray}

Replacing $\psi_{6c} \approx \Big[1- N\Big] \tilde{\psi}_{6c}$ into eqn. (A7), we obtain an equation for the conduction band electrons as follows,
$
(H + \delta H) \tilde{\psi}_{6c} = \tilde{E} \tilde{\psi}_{6c}, $
where the bare Hamiltonian $H_{0}$ is defined by 
\begin{eqnarray}
H &=& \frac{\hbar^{2} k^{2}}{2m}+ V + q\lambda \vec{\sigma} \cdot (\vec{k} \times \vec{E}) + \frac{q\lambda}{2} \vec{\nabla}. \vec{E},
\nonumber\end{eqnarray}
and $\delta H$ is given by
\begin{eqnarray}
\delta H &=& \frac{P^{2}}{3}\Big( \frac{2}{E_{G}} + \frac{1}{E_{G}+\Delta_{0}} \Big) k^{2}  \nonumber\\
&&-\frac{P^{2}}{3} \Big( \frac{1}{E_{G}} - \frac{1}{E_{G}+\Delta_{0}} \Big)\frac{e}{\hbar} i \vec{\sigma} \cdot  (\vec{k} \times \vec{k}) \nonumber\\
&& +\frac{eP^{2}}{3} \Big( \frac{1}{E_{G}^{2}} - \frac{1}{(E_{G}+\Delta_{0})^{2}}\Big) \vec{\sigma} \cdot (\vec{k} \times \vec{E} )  \nonumber\\
&& -\frac{eP^{2}}{6} \Big( \frac{2}{E_{G}^{2}} + \frac{1}{(E_{G}+\Delta_{0})^{2}}   \Big)\vec{\nabla}. \vec{E},     \label{tH'}
\nonumber\end{eqnarray}
with $E=(-1/e)\nabla V,$ 
where, we have used the following relations \cite{winkler},
\begin{eqnarray}
&&(\vec{\sigma} \cdot \vec{k})(\vec{\sigma} \cdot k) = k^{2}  +i \vec{\sigma} \cdot (\vec{k} \times \vec{k}), \nonumber\\
&& 9(\vec{T} \cdot \vec{k})(\vec{T}^{\dagger} \cdot \vec{k} ) = 2 k^{2}  -i \vec{\sigma} \cdot (\vec{k} \times \vec{k}).
\nonumber\end{eqnarray}

Thus, the total Hamiltonian for the conduction band: $H^{~'} = H + \delta H$ reads as
\begin{eqnarray}
H_{tot} &=& \frac{\hbar^{2} k^{2}}{2m^{*}} + qV  + q(\lambda +\delta \lambda)\vec{\sigma} \cdot ( \vec{k} \times \vec{E}) + (1 + \frac{\delta g}{2})\mu_{B}\vec{\sigma} . \vec{B} \nonumber\\
&&+  \frac{q}{2}(\lambda + \delta \lambda_{ D}) \vec{\nabla}. \vec{E}. \label{r-H}
\nonumber\end{eqnarray}
If we are dealing constant electric field, the last term cancels out and we finally arrive at Hamiltonian (3) of this manuscript.
Here, 
the effective mass $m^{*}$ and Kane model parameters $\delta g$, $\delta \lambda_{ S}$, and $\delta \lambda_{ D}$ are given by 
\begin{eqnarray}
\frac{1}{m^{*}} &=& \frac{1}{m} + \frac{2P^{2}}{3\hbar^{2}}\Big( \frac{2}{E_{G}} + \frac{1}{E_{G}+\Delta_{0}} \Big)\nonumber\\
\delta g &=&  -\frac{4m}{\hbar^{2}} \frac{P^{2}}{3} \Big( \frac{1}{E_{G}} - \frac{1}{E_{G}+\Delta_{0}} \Big),\label{r-g}\nonumber\\
\delta \lambda &=&  - \frac{P^{2}}{3} \Big( \frac{1}{E_{G}^{2}} - \frac{1}{(E_{G}+\Delta_{0})^{2}}\Big),\label{r-so}\nonumber\\
\delta \lambda_{D} &=& \frac{P^{2}}{3} \Big( \frac{2}{E_{G}^{2}} + \frac{1}{(E_{G}+\Delta_{0})^{2}} \Big).\label{r-da}
\nonumber\end{eqnarray} 
These renormalization factors $\delta g$, $\delta \lambda$, and $\delta \lambda_{ D}$ are the Kane model parameters in  the presence of the electric and magnetic fields \cite{winkler,kane}. 

According to the conventional $\vec{k} \cdot \vec{p}$ method, the renormalized Zeeman term is 
$
\frac{(g_{0}+\delta g)}2 \mu_{B} \vec{\sigma} \cdot \vec{B}
,$
where, $g_{0}=2$ is the bare $g$ factor. 
The renormalized $g$ factors, $g_{0}+\delta g$ of the Zincblende-type semiconductors have already been studied theoretically and experimentally \cite{kane}. 

The function L(T) in eqn(9) can be written as 
\beq  L(T)= \mu_{B}\left(1 -\frac{2m}{\hbar^{2}}\frac{1}{3a^{2}\hbar^{2}(1+ \alpha_{th}(T - 300))^{2}}\left(\frac{1}{E_{G}(0) - \frac{\alpha T^{2}}{T+\beta}} - \frac{1}{E_{G}(0) - \frac{\alpha T^{2}}{T+\beta} + \triangle_{0}}\right) \right).\nonumber~~~~~~~~~~~~~~~~~~~~~~~~~(A8)   \eeq
In eqn. (23), \beq \xi(T) = \left[\lambda + \frac{1}{3a^{2}\hbar^{2}(1+ \alpha_{th}(T - 300))^{2}}\left(\frac{1}{\left(E_{G}(0) - \frac{\alpha T^{2}}{T+\beta}\right)^{2}} - \frac{1}{\left(E_{G}(0) - \frac{\alpha T^{2}}{T+\beta} + \triangle_{0}\right)^{2}}\right) \right].\nonumber~~~~~~~~~~~~~~~~~~~~~~~~~(A9)  \eeq 
In eqn. (37) 
\begin{eqnarray}
R(T) &=&\left[\lambda + \frac{1}{3a^{2}\hbar^{2}(1+ \alpha_{th}(T - 300))^{2}}\left(\frac{1}{\left(E_{G}(0) - \frac{\alpha T^{2}}{T+\beta}\right)^{2}} - \frac{1}{\left(E_{G}(0) - \frac{\alpha T^{2}}{T+\beta} + \triangle_{0} \right)^{2}}\right) \right]\nonumber\\
&&\left(\frac{1}{m} + \frac{2}{3a^{2}\hbar^{2}(1+ \alpha_{th}(T - 300))^{2}}\left(\frac{2}{E_{G}(0) - \frac{\alpha T^{2}}{T+\beta}} + \frac{1}{E_{G}(0) - \frac{\alpha T^{2}}{T+\beta} + \triangle_{0}}\right)\right).\nonumber~~~~~~~~~~~~~~~~~~~~~~~~~(A10)   
\end{eqnarray}

\end{document}